# S-Parameter Uncertainties in Network Analyzer Measurements with Application to Antenna Patterns

*Nikolitsa YANNOPOULOU, Petros ZIMOURTOPOULOS*

Antennas Research Group, Dept. of Electrical Engineering and Computer Engineering, Democritus University of Thrace, V. Sofias 12, P.O. Box 63, 671 00 Xanthi, Greece

yin@antennas.gr, pez@antennas.gr, www.antennas.gr

**Abstract.** *An analytical method was developed, to estimate uncertainties in full two-port Vector Network Analyzer measurements, using total differentials of S-parameters. System error uncertainties were also estimated from total differentials involving two triples of standards, in the Direct Through connection case. Standard load uncertainties and measurement inaccuracies were represented by independent differentials. Complex uncertainty in any quantity, differentiably dependent on S-parameters, is estimated by the corresponding Differential Error Region. Real uncertainties, rectangular and polar, are estimated by the orthogonal parallelogram and annular sector circumscribed about the Differential Error Region, respectively. From the user's point of view, manufactures' data may be used to set the independent differentials and apply the method. Demonstration results include: (1) System error differentials for Short, matching Load and Open pairs of opposite sex standards; (2) System error uncertainties for VNA extended by two lengthy transmission lines of opposite sex end-connectors; (3) High uncertainties in Z-parameters against frequency of an appropriately designed, DC resistive, T-Network; (4) Moderate uncertainties in amplitude and phase patterns of a designed UHF radial discone antenna (azimuthally rotated by a built positioner, under developed software control of a built hardware controller) polarization coupled with a constructed gain standard antenna (stationary) into an anechoic chamber.*

## 1. Introduction

It is well known that, in full two-port Vector Network Analyzer (VNA) measurements, the parameters $S_{ij}$ ($i=1,2$; $j=1,2$) of a two-port Device Under Test (DUT), in general, result from their measurements $m_{ij}$ and the quantities $D$, $M$, $R$, $L$, $T$, $X$ and $D'$, $M'$, $R'$, $L'$, $T'$, $X'$, in the forward and reverse direction, respectively [1]. All the involved, implicitly dependent on the frequency, quantities are shown in the flow-graphs of Fig.1. $S_{11}$, $S_{21}$ are given by

$$S_{11} = \{[(m_{11} - D)/R][1 + (m_{22} - D')M'/R'] - L(m_{21} - X)(m_{12} - X')/(TT')\}/H, \quad (1)$$

$$S_{21} = \{[1 + (m_{22} - D')(M' - L)/R'](m_{21} - X)/T\}/H, \quad (2)$$

$$H = [1 + (m_{11} - D)M/R][1 + (m_{22} - D')M'/R'] - LL'(m_{21} - X)(m_{12} - X')/(TT') \quad (3)$$

as shown in Fig. 1(a). $S_{22}$, $S_{12}$ have similar expressions that result from (1), (2) by substituting the indices *1*, *2* with *2*, *1*, the symbols *D*, *M*, *R*, *L*, *T*, *X* with *D'*, *M'*, *R'*, *T'*, *L'*, *X'*, and vice-versa, as shown in Fig. 1(b).

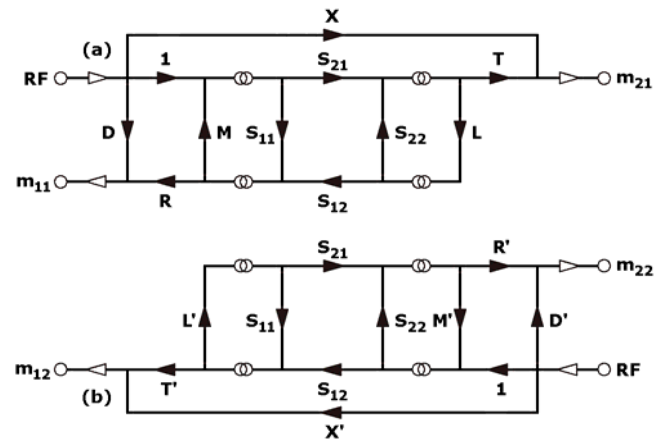

**Fig. 1.** Full two-port error model: (a) Forward, (b) Reverse.

These 12 quantities have been defined as system errors *Es* and a physical meaning has been given to them [2]. Accordingly, *D* is the directivity error $E_D$, *M* is the source match error $E_M$, *R* is the frequency response error $E_R$, *L* is the load match error $E_L$, *T* is the transmission tracking error $E_T$ and *X* is the isolation or crosstalk error $E_{X_S}$, in the forward direction, while the corresponding primed symbols have identical meaning, in the reverse direction. Although it is possible to define these errors in terms of elementary circuit quantities by setting forth a number of justifiable flow-graph simplification assumptions, as it has been shown by the authors for a typical VNA measurement system such as the one was used to get the results in this paper, this analysis is too extensive to be reproduced here.

Stumper, in 2003, gave non-generalised expressions for partial deviations in $S$-parameters in two-port VNA measurements due to calibration standard uncertainties only [3]. Total differential errors, including both standard uncertainties and measurement inaccuracies, have been given recently for full one-port VNA measurements, but these are also not generalised to full two-port VNA measurements [4]. To the best of the authors' knowledge, there are no analytical expressions for $S$-parameter total differentials to express uncertainty estimation in $S$-parameters resulting from full two-port VNA measurements of a DUT.

## 2. Theory

By applying the differential operator on (1)-(3) the total differential errors of the $S$-parameters, in the forward direction, were initially developed as

$$dS_{11} = \{TT'(1 - MS_{11})[R' + M'(m_{22} - D')](m_{11} - dD)$$
$$- RR'L(1 - L'S_{11})(m_{21} - X)(dm_{12} - dX')$$
$$- RR'L(1 - L'S_{11})(m_{12} - X')(dm_{21} - dX)]$$
$$+ M'TT'[(m_{11} - D)(1 - MS_{11}) - RS_{11}](dm_{22} - dD')$$
$$- TT'S_{11}(m_{11} - D)[R' + M'(m_{22} - D')]dM$$
$$+ TT'(m_{22} - D')[(m_{11} - D)(1 - MS_{11}) - RS_{11}]dM'$$
$$- (R'L(1 - L'S_{11})(m_{12} - X')(m_{21} - X)$$
$$+ TT'S_{11}[R' + M'(m_{22} - D')] )dR$$
$$- (RL(1 - L'S_{11})(m_{12} - X')(m_{21} - X)$$
$$- TT'[(m_{11} - D)(1 - MS_{11}) - RS_{11}] )dR'$$
$$- RR'(m_{12} - X')(m_{21} - X)[(1 - L'S_{11})dL - LS_{11}dL']$$
$$+ [(m_{11} - D)(1 - MS_{11}) - RS_{11}]$$
$$\cdot [R' + M'(m_{22} - D')](T'dT + TdT')\}/P, \qquad (4)$$

$$dS_{21} = \{-MTT'S_{21}[R' + M'(m_{22} - D')](dm_{11} - dD)$$
$$+ RR'LL'S_{21}(m_{21} - X)(dm_{12} - dX')$$
$$+ R[T'[R' + (m_{22} - D')(M' - L)]$$
$$+ R'LL'S_{21}(m_{21} - X')](dm_{21} - dX)$$
$$+ T'(R(m_{21} - X)(M' - L)$$
$$- M'TS_{21}[R + M(m_{11} - D)] )(dm_{22} - dD')$$
$$- TT'S_{21}(m_{11} - D)[R' + M'(m_{22} - D')]dM$$
$$+ T'(m_{22} - D')(R(m_{21} - X)$$
$$- TS_{21}[R + M(m_{11} - D)] )dM'$$
$$+ ( (m_{21} - X)(T'(m_{22} - D')(M' - L)$$
$$+ R'[T' + LL'S_{21}(m_{12} - X')] )$$
$$- TT'S_{21}[R' + M'(m_{22} - D')] )dR$$
$$+ (R(m_{21} - X)[T' + LL'S_{21}(m_{12} - X')]$$

$$- TT'S_{21}[R + M(m_{11} - D)] )dR'$$
$$+ R(m_{21} - X)[R'L'S_{21}(m_{12} - X') - T'(m_{22} - D')]dL$$
$$+ RR'LS_{21}(m_{12} - X')(m_{21} - X)dL'$$
$$- T'S_{21}[R + M(m_{11} - D)][R' + M'(m_{22} - D')]dT$$
$$+ (R(m_{21} - X)[R' + (m_{22} - D')(M' - L)] - TS_{21}$$
$$\cdot [R + M(m_{11} - D)][R' + M'(m_{22} - D')]dT'\}/P \qquad (5)$$

where

$$P = TT'[R' + M'(m_{22} - D')][R + M(m_{11} - D)]$$
$$- RR'LL'(m_{12} - X')(m_{21} - X). \qquad (6)$$

$dS_{22}$, $dS_{12}$ resulted from (4)-(5) by similar expressions.

The system error differentials were determined by the following procedure:

(i) According to full one-port VNA total differential errors [4] $dD$, $dM$, $dR$ were expressed in terms of differentials for standard uncertainties $dA$, $dB$, $dC$ and for their measurement inaccuracies $da$, $db$, $dc$. $dD'$, $dM'$, $dR'$ resulted by similar expressions.

(ii) $dX$ and $dX'$ resulted immediately as the differentials of the corresponding isolation measurements.

(iii) By applying the differential operator to the Direct Through connection expressions

$$L = (t_{11} - D) / [M(t_{11} - D) + R], \qquad (7)$$

$$T = R(t_{21} - X) / [M(t_{11} - D) + R] \qquad (8)$$

$dL$, $dT$ were initially developed as:

$$dL = [(1 - ML)(dt_{11} - dD) - LdR - (t_{11} - D)LdM]$$
$$/ [M(t_{11} - D) + R], \qquad (9)$$

$$dT = [-MT(dt_{11} - dD) + R(dt_{21} - dX)$$
$$+ (t_{21} - X - T)dR - (t_{11} - D)TdM]$$
$$/ [M(t_{11} - D) + R]. \qquad (10)$$

$L'$, $T'$ and $dL'$, $dT'$ were resulted from (7)-(10), by similar expressions.

Finally, by using the above expressions for system errors and their differentials (i) and (ii), L and T were appropriately expressed by

$$L = [(ab + ct_{11})C(B - A) + (bc + at_{11})A(C - B)$$
$$+ (ca + bt_{11})B(A - C)]/\!E$$
$$= [\Sigma (ab + ct_{11})C(B - A)]/\!E, \qquad (11)$$

$T = (t_{21} - X)$

$\qquad \cdot [(A - B)(a - b)(B - C)(b - c)(C - A)(c - a)]Æ F$

$\qquad = (t_{21} - X) \ [\prod (A - B)(a - b)]Æ F, \qquad (12)$

$E \equiv (ab + ct_{11})(B - A) + (bc + at_{11})(C - B)$

$\qquad + (ca + bt_{11})(A - C)]$

$\qquad = \sum (ab + ct_{11})(B - A), \qquad (13)$

$F \equiv cC(B - A) + aA(C - B) + bB(A - C)$

$\qquad = \sum cC(B - A) \qquad (14)$

where $\sum$ and $\prod$ produce two more terms, from the given one, by cyclic rotation of the letters $a$, $b$, $c$ and $A$, $B$, $C$ respectively. $L'$, $T'$ are derived from (11)-(14) by similar expressions.

Thus $dL$, $dT$ were finally expressed as

$dL = \{\sum (B - C)(b - t_{11})(c - t_{11})$

$\qquad \cdot [(B - C)(b - a)(c - a)dA$

$\qquad - (b - c)(B - A)(C - A)da]$

$\qquad + [\prod (A - B)(a - b)]dt_{11}\}Æ E^2$

$\qquad = (1/Æ^2)\{(B - C)(b - t_{11})(c - t_{11})$

$\qquad \cdot [(B - C)(b - a)(c - a)dA$

$\qquad - (b - c)(B - A)(C - A)da]$

$\qquad + (C - A)(c - t_{11})(a - t_{11})$

$\qquad \cdot [(C - A)(c - b)(a - b)dB$

$\qquad - (c - a)(C - B)(A - B)db]$

$\qquad + (A - B)(a - t_{11})(b - t_{11})$

$\qquad \cdot [(A - B)(a - c)(b - c)dC$

$\qquad - (a - b)(A - C)(B - C)dc]$

$\qquad + [(A - B)(a - b)(B - C)(b - c)$

$\qquad \cdot (C - A)(c - a)]dt_{11}\}, \qquad (15)$

$dT = (t_{21} - X)\{\sum (b - c)(B - C)$

$\qquad \cdot [(t_{11} - c)(b - a)^2 B(A^2 + C^2)$

$\qquad + (b - t_{11})(c - a)^2 C(A^2 + B^2)$

$\qquad - 2ABC(b - c)(t_{11}(b + c - 2a) - bc + a^2)]$

$\qquad \cdot [(B - C)(b - a)(c - a)dA$

$\qquad - (b - c)(B - A)(C - A)da]/(E^2 F^2)$

$\qquad + [\prod (A - B)(a - b)]$

$\qquad \cdot \{[(t_{21} - X)\sum a(B - C)Æ E]dt_{11} + dt_{21} - dX\}/(EF)$

$\qquad = (1/(E^2 F^2))(t_{21} - X)\{(b - c)(B - C)$

$\qquad \cdot [(t_{11} - c)(b - a)^2 B(A^2 + C^2)$

$\qquad + (b - t_{11})(c - a)^2 C(A^2 + B^2)$

$\qquad - 2ABC(b - c)(t_{11}(b + c - 2a) - bc + a^2)]$

$\qquad \cdot [(B - C)(b - a)(c - a)dA$

$\qquad - (b - c)(B - A)(C - A)da]$

$\qquad + (c - a)(C - A)$

$\qquad \cdot [(t_{11} - a)(c - b)^2 C(B^2 + A^2)$

$\qquad + (c - t_{11})(a - b)^2 A(B^2 + C^2)$

$\qquad - 2ABC(c - a)(t_{11}(c + a - 2b) - ca + b^2)]$

$\qquad \cdot [(C - A)(c - b)(a - b)dB$

$\qquad - (c - a)(C - B)(A - B)db]$

$\qquad + (a - b)(A - B)$

$\qquad \cdot [(t_{11} - b)(a - c)^2 A(C^2 + B^2)$

$\qquad + (a - t_{11})(b - c)^2 B(C^2 + A^2)$

$\qquad - 2ABC(a - b)(t_{11}(a + b - 2c) - ab + c^2)]$

$\qquad \cdot [(A - B)(a - c)(b - c)dC$

$\qquad - (a - b)(A - C)(B - C)dc]\}$

$\qquad + (A - B)(a - b)(B - C)(b - c)(C - A)(c - a)/(EF)$

$\qquad \cdot \{(1/E)(t_{21} - X)$

$\qquad \cdot [a(B - C) + b(C - A) + c(A - B)]dt_{11}$

$\qquad + dt_{21} - dX\} \qquad (16)$

while $dL'$, $dT'$ resulted from (15), (16) respectively by similar expressions.

All of the above expressions were mechanically cross-verified using a developed software program for symbolic computations.

The set of these 12 complex differentials, considered as functions of frequency, can be possibly called "the differential error core of the full two-port VNA measurement system".

Therefore, each $S$-parameter has finally a total differential error $dS$ expressed in terms of 22 independent complex differentials: 6 to express the standard load uncertainties $dA$, $dB$, $dC$, $dA'$, $dB'$, $dC'$ and 16 to express the measurement inaccuracies $dm_{ij}$, $da$, $db$, $dc$, $da'$, $db'$, $dc'$, $dX$, $dX'$, $dt_{ij}$, that is in terms of 88 independent real quantities.

In this way, any complex quantity, differentially dependent on one or more $S$-parameters, has an expressible total differential. For example, the $S$-to-$Z$ parameter general relations, as they developed by Beatty and Kerns in 1964 [5] and corrected by themselves [6] (although an additional correction is needed, when $Z_0$ is different for each port) were used to develop $dZ$ differentials as follows

$$dZ_{11} = 2Z_0[(1 - S_{22})^2 dS_{11} + (1 - S_{22})S_{21}dS_{12}$$

$$+ (1 - S_{22})S_{12}dS_{21} + S_{12}S_{21}dS_{22}]$$

$$/ [(1 - S_{11})(1 - S_{22}) - S_{12}S_{21}]^2, \qquad (17)$$

$$dZ_{21} = 2Z_0[(1 - S_{22})S_{21}dS_{11} + S_{21}^2 dS_{12}$$

$$+ (1 - S_{11})(1 - S_{22})dS_{21} + (1 - S_{11})S_{21}dS_{22}]$$

$$/ [(1 - S_{11})(1 - S_{22}) - S_{12}S_{21}]^2 \qquad (18)$$

while, $dZ_{22}$, $dZ_{12}$ result by similar expressions.

Since any total differential is a linear combination of independent partial differentials, analytic geometry can be used to represent a complex differential term by a corresponding total Differential Error Region (DER) on the complex plane. A total DER can be exactly illustrated from its, analytically described, independent DERs.

Complex uncertainty in a quantity can be estimated by its exact DER. For example, if the independent terms have a rectangular parallelogram DER or a circular about the origin DER, then the uncertainty in any $S$-parameter is estimated by a convex DER. In particular, if standard matching Loads are used, then the $S$-DER contour is a piecewise curve composed of $2[4(N-2)] = 160$ line segments and circular arcs, at most, while if matching Loads are not used, then the contour is a closed polygonal line with $4N = 88$ vertices, at most.

Real uncertainty in a quantity can not be directly estimated from a complex differential. This is due to the fact that the derivative of a non-constant real function in complex variable does not exist. Therefore, to estimate uncertainty in real and imaginary parts of a given quantity, we use the rectangular projections of the orthogonal parallelogram circumscribed about the DER and thus the notion of the Differential Error Intervals (DEIs) was defined [4]. Sharp inequalities define the end-points of each DEI. Respectively, uncertainty estimation in magnitude (amplitude) and argument (phase) can be calculated from sharp inequalities, for radius and angle, which are defined exactly by the Differential Error Radiuses and Differential Error Angles of the annular sector circumscribed about the DER.

# 3. Results

## 3.1 SLOdT System Error Differentials

The commonly used Short, matching Load and Open calibration standards, in pairs of opposite sex connectors for the direct Through connection case (SLOdT), with

$$A = A' = -1,$$

$$B = B' = 0,$$

$$C = C' = 1$$

were substituted in (11)-(16) to produce the following expressions for the corresponding system errors and their differentials

$$L = (a - c)(b - t_{11})$$

$$/ [t_{11}(a + c - 2b) + b(a + c) - 2ca], \qquad (19)$$

$$T = 2(t_{21} - X)(a - b)(b - c)$$

$$/ [t_{11}(a + c - 2b) + b(a + c) - 2ca], \qquad (20)$$

$$dL = \{(b - t_{11})(c - t_{11})[(b - a)(c - a)dA + 2(b - c)da]$$

$$+ 2(c - t_{11})(a - t_{11})[2(c - b)(a - b)dB + (c - a)db]$$

$$+ (a - t_{11})(b - t_{11})[(a - c)(b - c)dC + 2(a - b)dc]$$

$$+ 2(a - b)(b - c)(c - a)dt_{11}\}$$

$$/ [t_{11}(a + c - 2b) + b(a + c) - 2ca]^2, \qquad (21)$$

$$dT = (t_{21} - X)\{(b - c)(b - t_{11})(c - a)^2$$

$$\cdot [(b - a)(c - a)dA + 2(b - c)da]$$

$$+ 2(c - a)[(t_{11} - a)(c - b)^2 - (c - t_{11})(a - b)^2]$$

$$\cdot [2(c - b)(a - b)dB + (c - a)db]$$

$$+ (a - b)(b - t_{11})(a - c)^2$$

$$\cdot [(a - c)(b - c)dC + 2(a - b)dc]\}$$

$$/ [t_{11}(a + c - 2b) + b(a + c) - 2ca]^2(c - a)^2$$

$$+ 2(a - b)(b - c)\{(t_{21} - X)(2b - a - c)dt_{11}$$

$$/ [t_{11}(a + c - 2b) + b(a + c) - 2ca]$$

$$+ dt_{21} - dX\}/[t_{11}(a + c - 2b) + b(a + c) - 2ca]. \qquad (22)$$

The remaining SOLdT system errors and their differentials have already been expressed either in section 2(ii), above, or in full one-port VNA measurements [4].

## 3.2 System Error Uncertainties

To apply the expressions of the previous sub-section 3.1 for a SLOdT measurement system, from the user's point of view, manufacturers' data can be used to set the independent differentials.

Thus, for our SLOdT measurement system, which has been described in details in [4], the following manufacturers' standard uncertainty data were considered to define the endpoints of the corresponding polar differential intervals

$$-0.01 \le d|A|, \ d|A'| \le 0,$$

$$-2° \le d\angle A, \ d\angle A' \le +2°,$$

$$0 \le |dB|, \ |dB'| \le 0.029,$$

$$-0.01 \le d|C|, \ d|C'| \le 0,$$

$$-2° \le d\angle C, \ d\angle C' \le +2°.$$

Since the measurements are usually given in decibels and degrees, the measurement inaccuracy consideration admits a differential representation of the form

$$dm/m = (ln10/20)d|m|_{decibels} + j(\pi/180) \ d\angle m_{degrees}.$$

The endpoints of the independent inaccuracy differentials were determined by our VNA internal consideration, which permits a 3-digit display, of either amplitude or phase, with a variable accuracy, as follows:

(i) By either ±0.01 decibels in amplitude, when its displayed value is approximately within ±8 decibels or ±0.1 decibels, otherwise, and

(ii) By either ±0.1 degrees in phase, if its displayed value is approximately within ±80 degrees or ±1 degrees, otherwise.

Notably, other, external to VNA, measurement considerations are possible as well. These include, for example, all the cases in which the measurements result from an additional application of a pre-processing technique on the collected VNA data.

Consequently, since each independent variable $z$ has a differential in the polar form $dz = e^{jy}(d|z| + j|z|dy)$, its corresponding DER is either an orthogonal parallelogram, rotated about the origin, or a circle, with its center at the origin. Therefore, each $S$-DER, in our SOLdT system, is a sum of 20 orthogonal parallelograms and 2 circles, with a contour of 160 vertices, at most.

Our $Z_0 = 50 \ \Omega$ VNA measurement system was extended into an anechoic chamber by two lengthy transmission lines. The end-connector of the shorter line was attached through a matching rotary transmission air-line piece to a matching right-angle adapter, the open end of which defines the reference plane of port 1, for the measurement system as a whole. The reference plane of the system port 2 is defined by the end connector of the longer line. The two ports are of opposite sex, to fulfil the most accurate zero-length connection requirement for the Direct Through case.

Measurements were made from 2 to 1289 MHz, in 13 MHz steps. The $L, T, X$ system errors, in the forward direction are shown in Fig. 2 against frequency. The illustrations of the $L', T', X'$ errors, in the reverse direction, are omitted, as similar.

The frequency $f = 1003 \ MHz$ was selected to reveal the details of $L, T, X$ DERs as shown in Fig. 3, where the contours have small circles as their vertices. Typical DER illustrations for the remaining system errors already have been given in the full one-port measurements [4].

All DERs were checked against computed differences placed over them, with satisfactory results in all cases, except for the $T$ and $T'$ DERs, where a rather appreciable divergence was observed. A closer examination proved that the considered manufacturer's uncertainty value $dB$ is responsible for this $T$-DER deviation.

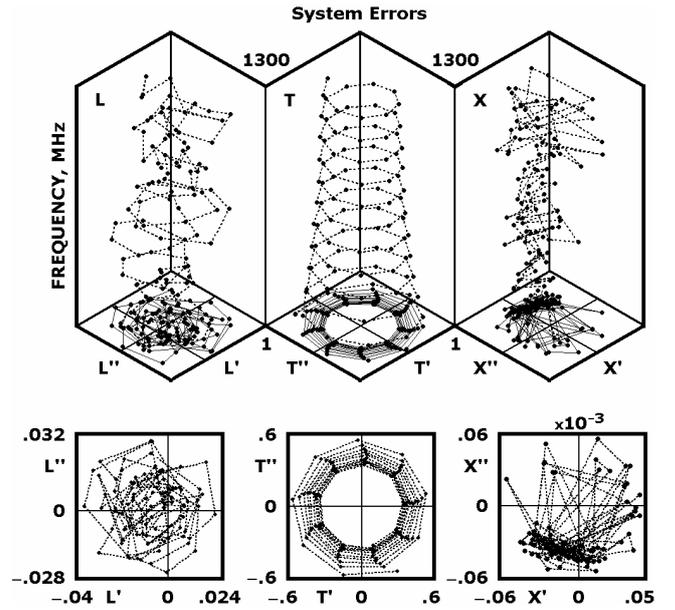

**Fig. 2.** $L, T$ and $X$ system errors.

To demonstrate this fact at the selected frequency, the partial differential $d_BT$ of $T$ with respect to $B$, and the corresponding partial difference $\delta_BT$, are shown in Fig. 3, for the given uncertainty $dB$. Since $\delta_BT/d_BT \rightarrow 1$, when $dB \rightarrow 0$, as (8), (20) and (22) verify that, to reveal further this type of behavior, pairs of $d_BT$ and $\delta_BT$ were illustrated by gradually reducing the given $dB$ down to $dB{:}\delta$, where a rather satisfactory coincidence is achieved, as shows the grey shade in Fig. 3. Similar conclusions hold for the $T'$-DER.

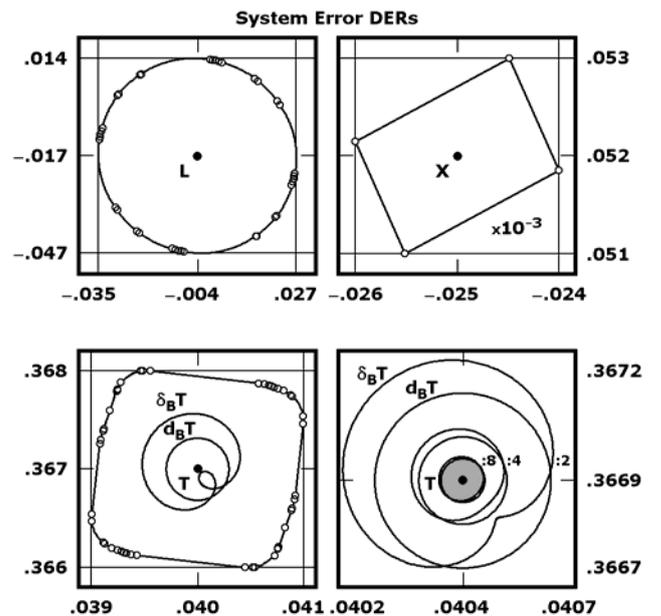

**Fig. 3.** $L, T$ and $X$ DERs, at 1003 MHz.

### 3.3 High Uncertainties in the Z-parameters of a DC-Resistive T-Network

An appropriately designed T-Network of common resistors, with combined nominal DC values $Z_1=24.2\ \Omega$, $Z_2=120\ \Omega$ for the horizontal arms and $Z_{12}=1.1\ \Omega$ for the vertical arm, were soldered on a couple of type-N base connectors of opposite sex and enclosed in an aluminium box, to form a two-port DUT for SLOdT measurements. As shown in the photography of Fig. 4, this DUT was placed between the ports of our SLOdT VNA system extended into the anechoic chamber.

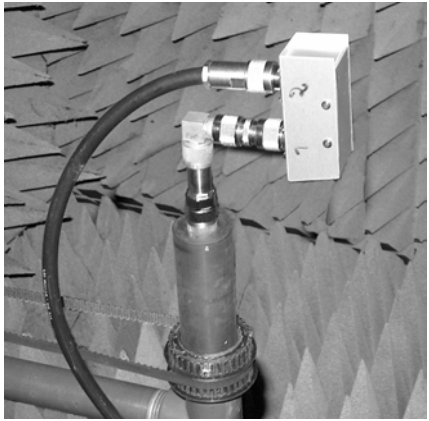

**Fig. 4.** The T-Network box in situ.

To illustrate the variation of S-parameter DERs for the T-Network against frequency, a number of selected S-DER frames are shown in Fig. 5, as beads on a space-curved filament.

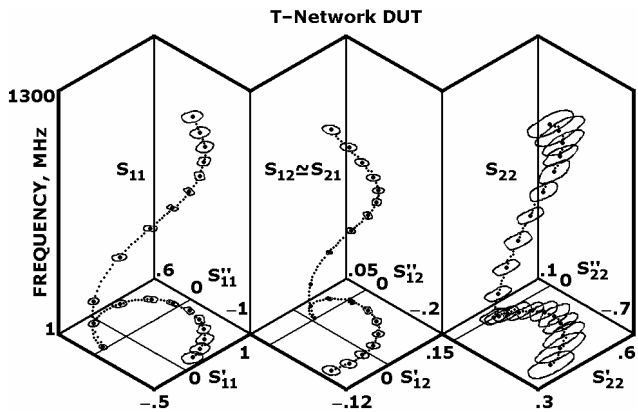

**Fig. 5.** S-parameter DERs against frequency.

Fig. 6 indicates that since the measurements $S_{12}$ and $S_{21}$ and their differential errors are almost overlapping, the reciprocity is approximately satisfied for the T-Network DUT. It is worth mentioning that $S_{11}$ and $S_{22}$ DERs were greater than those resulted from appropriately organised full one-port measurements for the T-Network.

High uncertainty estimation values in resistance $R$ and reactance $X$ parts of $Z$-parameters against frequency are observed in Fig. 7, where the solid lines mark their DC values.

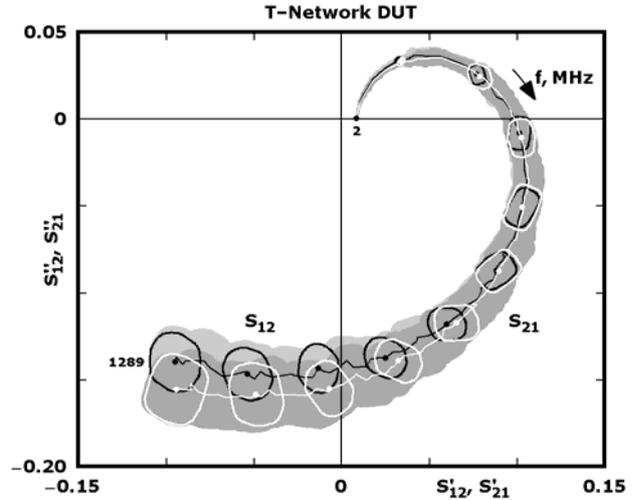

**Fig. 6.** $S_{12}$ and $S_{21}$ against frequency.

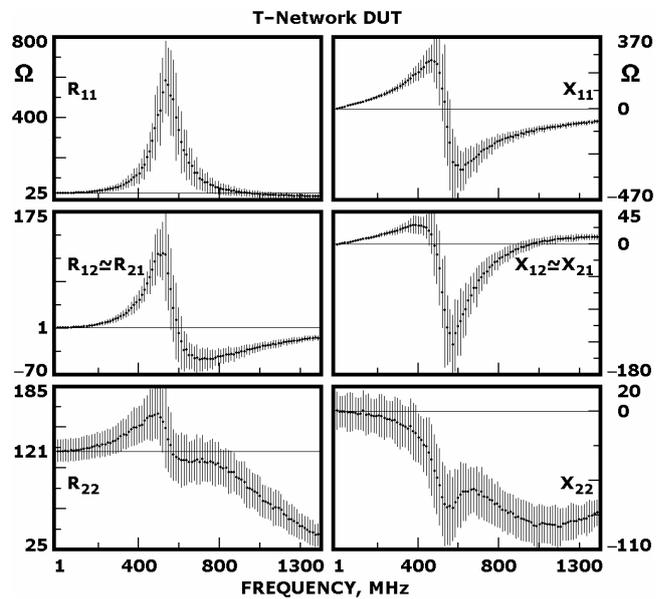

**Fig. 7.** $Z$ DEIs against frequency.

### 3.4 Moderate Uncertainties in the Patterns of a UHF Radial Discone Antenna

A designed and built UHF radial discone antenna into an anechoic chamber was used as Antenna Under Test (AUT). The design of AUT was accomplished by a developed suite of software simulation visual tools [4]. The considered model, the coordinate system, the feeding and the built details are shown in Fig 8, 8a, 8b and 8c, respectively. The predicted radiation pattern is almost z-axis symmetrical and almost $\theta$-polarized. The AUT was coupled with a gain standard antenna. Fig. 8d shows the test arrangement, where AUT is connected at port 1 and gain standard at port 2, of the extended measurement system that is described in 3.2. A radial disk was built by 8 wires of 8.1 cm in length, soldered on a tip of an N-type male connector. A radial frustum cone of a 120° flare angle was built by 8 wires of 12.5 cm in length, soldered on a circular wired ring of 1.6 cm in diameter. Bare copper wire of $\varnothing1.5$ mm was used.

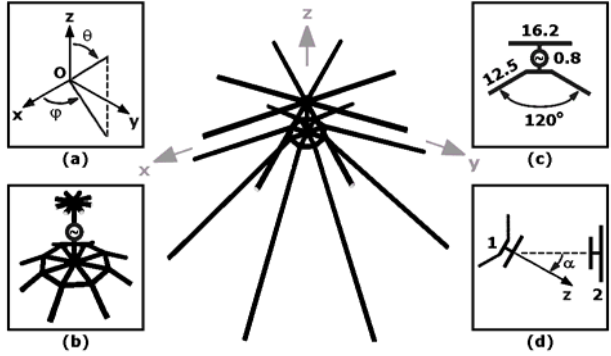

**Fig. 8.** UHF Radial Discone AUT simulation model.

The disk and cone AUT parts are shown in Fig. 9, where the two nuts were used to fix the cone on the port 1 type-N female connector. In order to ensure a well-defined reference plane for the AUT, to match as exactly as possible with that of the port 1 connector, the AUT was assembled by these two separate parts, as shown in Fig. 10.

The AUT was azimuthally rotated by a 360° built positioner, under the developed software control of a built hardware controller. The stationary antenna was the UHF gain antenna standard shown in Fig. 11 [7], [8]. As shown in Fig. 8d, azimuth $\alpha$ is corresponded to direction $(\theta, \varphi)$ by

$0° \leq \alpha < 180° \leftrightarrow (\theta = \alpha, \varphi), 0° \leq \varphi < 180°,$

$180° \leq \alpha < 360° \leftrightarrow (\theta = 360° - \alpha, \varphi + 180°); \varphi = 0°.$

Under a number of appropriate conditions for the separation of antennas, their reflection coefficients and polarizations, the $S$-parameters of the DUT can be used to express the characteristics of the AUT [9] and to calculate its gain [10]. In particular, $S_{12}$ and $S_{21}$ are analogous to the receiving and transmitting AUT patterns, respectively. These two $S$-parameters are equal, if reciprocity holds. In any case, we can consider the complex AUT radiation pattern $P$ to be analogous to the mean value $S$ of $S_{12}$ and $S_{21}$, that is $P = kS$, where $k$ is a constant.

We present results for the amplitude and phase radiation patterns, referenced to the maximum $|P|$. Thus, we initially considered the $P_{max} \equiv (k|S|_{max}) \angle S_{max}$, that is one corresponding to $max\{|S|: 0° \leq \alpha < 360°\}$, and then we expressed the normalized radiation pattern by $E = P/P_{max}$. Therefore, the amplitude and phase of $E$, and their differentials are given by

$|E| = |S| / |S|_{max},$        (23)

$\angle E = \angle S - \angle S_{max},$        (24)

$d|E| = d|S| / |S|_{max} - (|S| / |S|_{max}^2)d|S|_{max},$        (25)

$d\angle E = d\angle S - d\angle S_{max}.$        (26)

The $S$-parameters were measured against azimuth, in steps of $(10/1480) \times 360°$, at 1250 MHz. The $S$-DERs at $\alpha = 319°$ of $S_{max} \leftrightarrow P_{max}$ at $(\theta = 41°, \varphi = 180°)$ are detailed illustrated in Fig. 12, along with all inaccuracies $di$ and all uncertainties $du$ contributing to them. Notably, the circumscribed annular sectors result to uncertainty estimations.

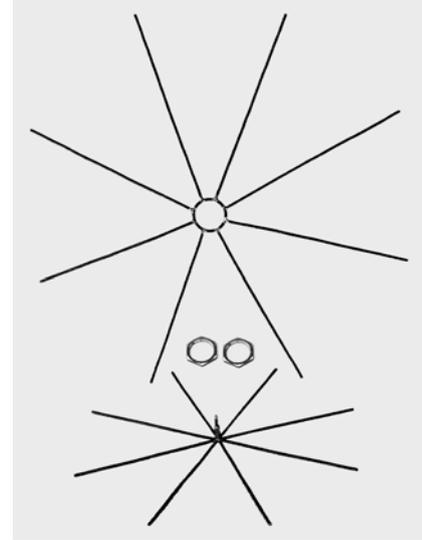

**Fig. 9.** The built parts of the Radial Discone AUT.

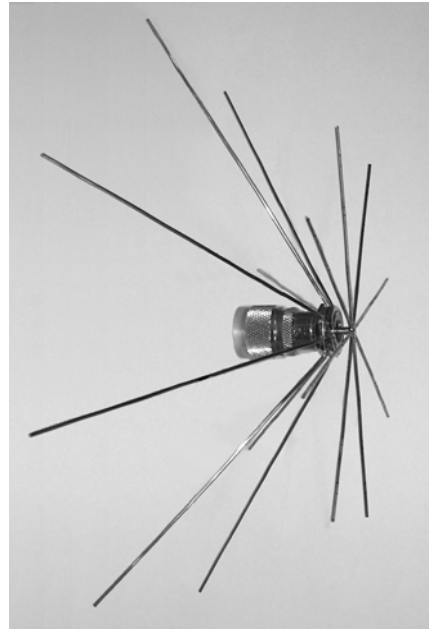

**Fig. 10.** The assembled AUT.

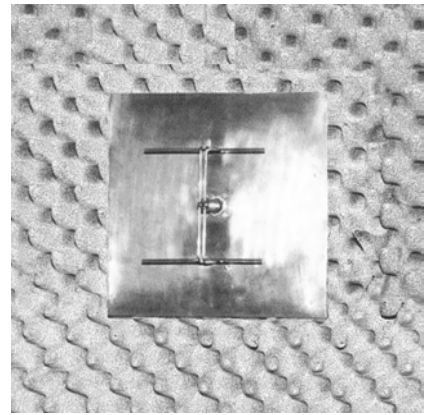

**Fig. 11.** The built UHF Gain Standard antenna.

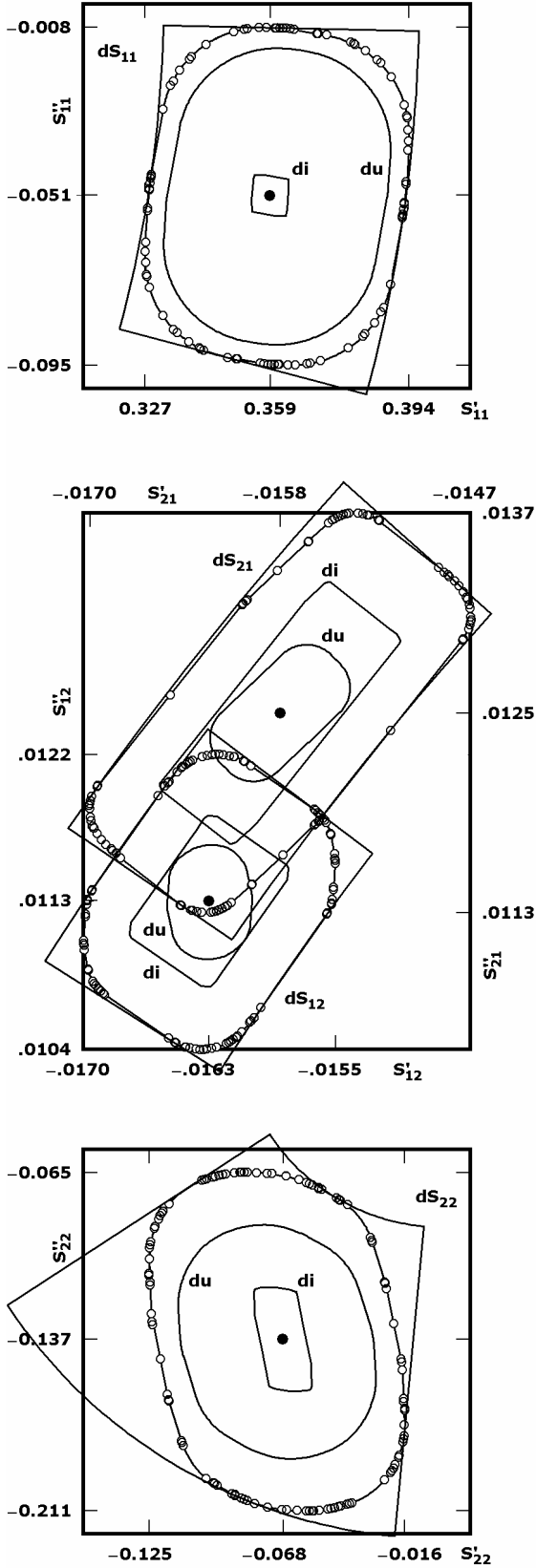

**Fig. 12.** *S-parameter DERs at 1250 MHz.*

The amplitude and phase uncertainties in the *S*-parameters are estimated by the real intervals, which are defined by the Differential Error Radiuses and Differential Error Angles that form the circumscribed annular sector about the *S*-DER. The *S*-parameter measurements on zOx plane, are illustrated in Fig. 13, against $\theta$. The almost identical $S_{12}$ and $S_{21}$ curves imply an approximate reciprocity. The grey strips result from the series of overlapping *S*-DERs, which illustrate the complex *S*-parameter uncertainties. For each of $S_{12}$ and $S_{21}$, there are two, almost coincide, grey strips, which result on $\varphi = 0\,°$ and $\varphi = 180\,°$ semiplanes, respectively. This fact is in accordance with the predicted symmetry of the AUT radiation pattern. While some variation exists in the $S_{11}$ and its complex uncertainty of the rotated AUT against azimuth, the $S_{22}$ of the stationary standard remains rather constant, as it was expected.

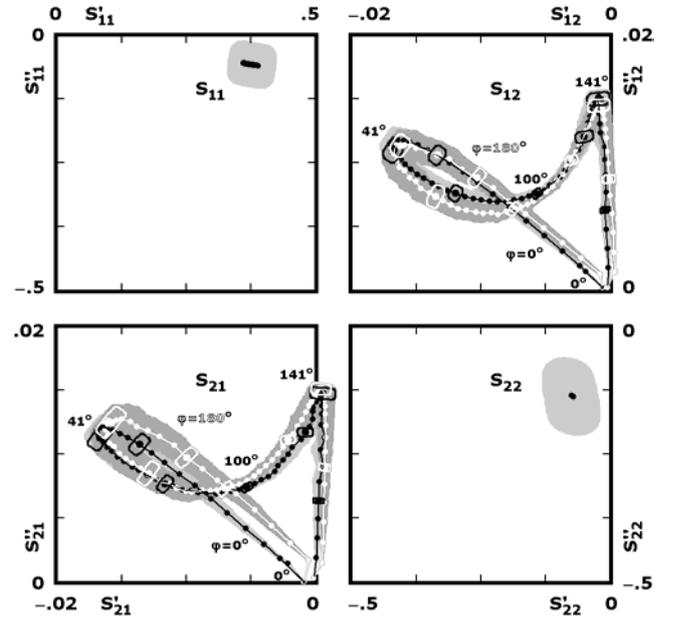

**Fig. 13.** *S-parameters and their DERs against θ, at 1250 MHz.*

The measured amplitude $|E|$ and phase $\angle E$ AUT normalized radiation patterns, along with their uncertainty estimations $\Delta|E|$ and $\Delta\angle E$, on zOx plane, at 1250 *MHz*, were computed by applying (23)-(26). The polar form of the results is illustrated at Fig. 14. The dashed lines are the predicted, by simulation, radiation patterns.

We defined the relative amplitude uncertainty by $\Delta_R|E| = (|E| \pm \Delta|E|) / |E|$ and we denoted by $\pm\Delta\angle E$, $\pm\Delta|E|$ and $\pm\Delta_R|E|$ the signed maximum values of the estimated uncertainties, respectively. These quantities, along with the ratio of $max|di|$ to $max|du|$ that contribute to *S*-DER, are shown in Fig. 15 against $\theta$, on the zOx plane. These, rather moderate, uncertainty values may be compared to the published maximum absolute uncertainty in our VNA measurements that are $\pm\Delta_R|E| = 0.3$ decibels and $\pm\Delta\angle E = 5\,°$, when equal maximum signal levels are applied to the input ports of its three receivers [11].

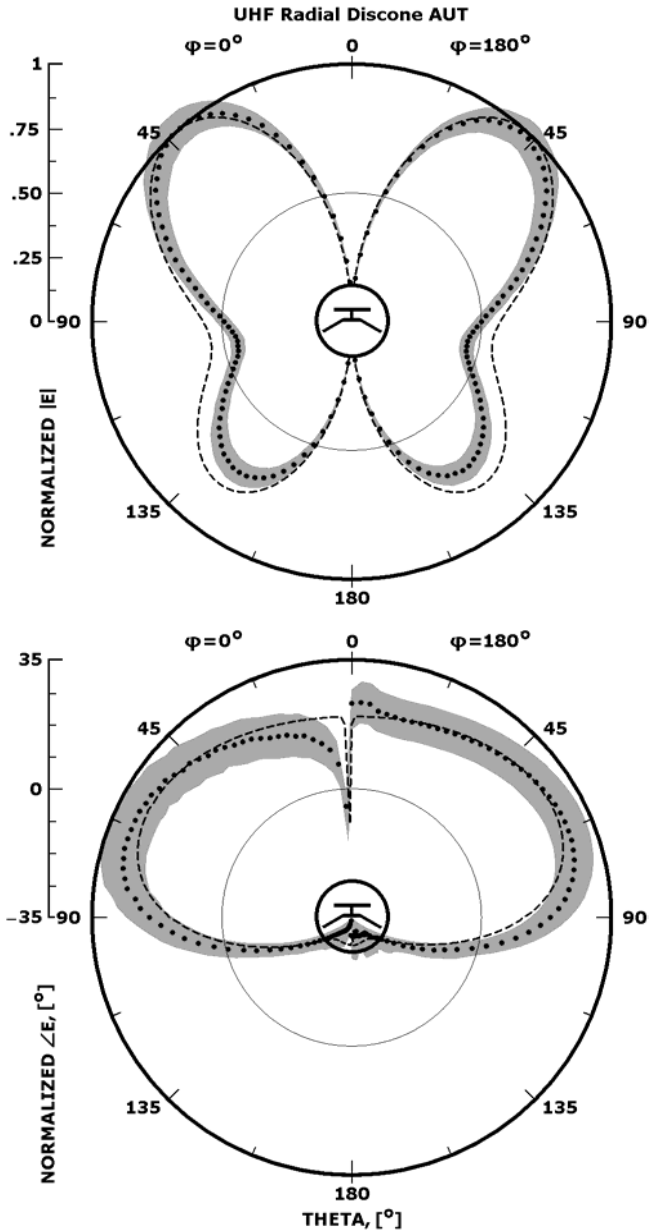

**Fig. 14.** AUT patterns on zOx plane, at 1250 MHz.

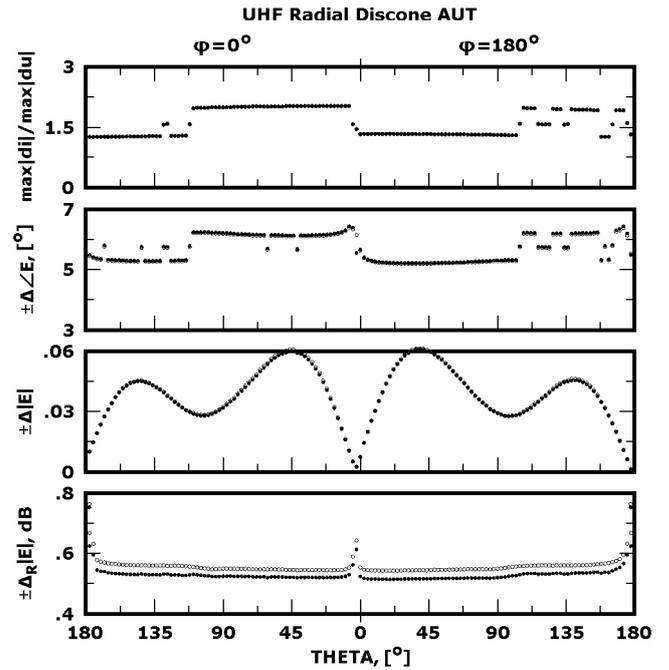

**Fig. 15.** Positive (●) and negative (○) maximum uncertainties in AUT patterns on zOx plane against θ, at 1250 MHz.

## About Authors...

**Nikolitsa YANNOPOULOU** was born in Chania, Crete, Greece in 1969. She graduated in 1992 from Electrical Engineering at Democritus University of Thrace, Xanthi, Greece and since then she is an unpaid Research Assistant with Antennas Research Group at Democritus University. She received the MSc degree with full marks in Microwaves at Democritus University in 2003. She completed her Dissertation on Antennas and she has to defend her PhD Thesis at the same University. Her research interests are in antennas: theory, design, software, simulation, construction, measurements and virtual laboratories.

**Petros ZIMOURTOPOULOS** is Assistant Professor in Electrical Engineering and Computer Engineering at Democritus University of Thrace.